\documentclass[letterpaper, 10 pt, conference]{ieeeconf}

\usepackage{amsfonts}

\begin{document}

\title{Quantum Error Correcting Subsystem Codes From Two Classical Linear Codes}
\author{\authorblockN{Dave Bacon}
\authorblockA{Department of Computer Science \& Engineering\\
University of Washington\\
Seattle, WA 98195 USA\\
Email: dabacon@cs.washington.edu} \and
\authorblockN{Andrea Casaccino}
\authorblockA{
Department of Information Engineering\\
University of Siena \\
Siena, Italy \\
 Email: ndr981@tin.it} } \maketitle
\begin{abstract}
The essential insight of quantum error correction was that quantum information
can be protected by suitably encoding this quantum information across multiple
independently erred quantum systems. Recently it was realized that, since the
most general method for encoding quantum information is to encode it into a
subsystem, there exists a novel form of quantum error correction beyond the
traditional quantum error correcting subspace codes.  These new quantum error
correcting subsystem codes differ from subspace codes in that their quantum
correcting routines can be considerably simpler than related subspace codes.
Here we present a class of quantum error correcting subsystem codes constructed
from two classical linear codes.  These codes are the subsystem versions of the
quantum error correcting subspace codes which are generalizations of Shor's
original quantum error correcting subspace codes. For every Shor-type code, the
codes we present give a considerable savings in the number of stabilizer
measurements needed in their error recovery routines.
\end{abstract}
\begin{keywords}
quantum error correction, subsystem codes, stabilizer codes
\end{keywords}

 \maketitle

Real quantum systems are open quantum systems which can couple in an unwanted
manner to an environment or control system and lose their intrinsic quantum
nature through the processes of decoherence, quantum noise, and imprecise
measurement, preparation, and control.  These problems serve as an obstacle
towards the eventual construction of a robust large scale quantum
computer\cite{Landauer:94a,Landauer:95a,Chuang:95b,Unruh:95a}. If left
unchecked, these problems turn a quantum computer into a classical information
processing device, or even worse, into a machine which can enact no computation
at all. Fortunately, however, soon after the discovery that quantum computers
were more efficient at solving certain problems than classical
computers\cite{Shor:94a,Grover:96a}, it was discovered that, under specific
reasonable physical assumptions, a fault-tolerant quantum computer could be
built. In particular, a set of {\em threshold theorems} for fault-tolerant
quantum computation were
established\cite{Aharonov:97a,Aharanov:99a,Knill:98a,Knill:98b,Kitaev:97b,Kitaev:97c}.
These theorems proved (or gave a heuristic proof) that if decoherence, quantum
noise, and lack of control were all small enough in comparison to the ability
to control the quantum system (below some threshold or thresholds), then these
noisy imprecise bare devices could be efficiently put together in a fashion
which decreased the failure probability of a quantum computer to any desired
level.

A central insight used in the theory of fault-tolerant quantum computation is
that quantum information can be encoded into what is known as a quantum error
correcting code\cite{Shor:95a,Steane:96a}.  Such codes spread quantum
information across many physical quantum bits (qubits) of a system and protect
the quantum information so encoded from the undesired effects which cause a
loss of quantum coherence.  In the original theory of quantum error correction,
quantum information was encoded into a subspace of the Hilbert space of many
quantum systems.  Such subspace codings, however, are not the most general way
to encode quantum information into a quantum system. The most general method to
encode quantum information is to encode it into a subsystem\cite{Knill:00a}.
This realization has recently led to the discovery of a new method for perform
quantum error correction using quantum error correcting
subsystems\cite{Kribs:05a,Kribs:05b,Kribs:06a,Nielsen:05a}. While every quantum
error correcting subsystem can be turned into a subspace code, subsystems codes
differ significantly in {\em how} quantum error correction is performed on the
encoded quantum information. Of particular significance is that quantum error
correcting subsystems can significantly reduce the number of stabilizer
measurements needed during their quantum error recovery
routines\cite{Poulin:05a,Bacon:06a,Klappenecker:06a}. This in turn can lead to
significantly improved thresholds for fault-tolerant quantum
computation\cite{Thaker:06a,Aliferis:06a}.

In this paper we construct a class of new quantum error correcting subsystem
codes.  These codes are a generalization of a code presented by one of us in
\cite{Bacon:06a} where they were put to use in an attempt to construct
self-correcting quantum memories.  The codes we describe are in the class of
stabilizer subsystem codes described by Poulin\cite{Poulin:05a}.  In particular
they can be described as the subsystem version of the class of codes arising
from generalizing Shor's original quantum error correcting subspace
codes\cite{Shor:95a}. The subsystems codes we describe are constructed by
taking two classical linear codes and, instead of concatenating them in the
manner of Shor, constructing a new quantum code which is a subsystem quantum
error correcting code directly from these classical linear codes.  In Shor's
original construction a bit flip error correcting code is concatenated with a
phase flip error correcting code (or vice versa.)  The order of these
concatenations presents an asymmetry in the recovery routine for these codes.
If, for example a bit flip code is used on the lowest level, then the recovery
procedure for the bit flip code must be enacted for every lowest level code,
whereas the phase flip code recovery routine need only be enacted on the next
level of the code.  In the codes we construct, this asymmetry between bit flips
and phase flips in the recovery routine is removed.  This leads to codes which
have substantially simpler error recovery routines, but provide the same amount
of protection as the generalized Shor-type codes.  Since the threshold for
fault-tolerant quantum computation depends significantly on the complexity of
the error recovery routine, these codes offer the hope of a substantial
improvement in the threshold for fault-tolerant quantum
computation\cite{Thaker:06a,Aliferis:06a}.


\section{Classical and Quantum Linear Codes}

We begin by briefly reviewing classical and quantum linear codes in order to
establish conventions for our future constructions.

We will work with classical codes over the finite field ${\mathbb F}_2$.
${\mathbb F}_2$ is the field with two elements $0$ and $1$ with addition
defined as $0+0=1+1=0$ and $0+1=1+0=1$ and multiplication $0 \cdot 0=0 \cdot
1=1 \cdot 0=0$ and $1 \cdot 1=1$.  Ordered $n$-tuples of ${\mathbb F}_2$ form a
linear space, $({\mathbb F}_2)^{n}$.  A classical binary linear code on $n$
bits is a subspace ${\mathcal C}$ of $({\mathbb F}_2)^n$.  A subspace can be
described by a set of $k$ basis vectors.  We can list such a set of $k$ basis
vectors in a $k \times n$ matrix, $G$, with elements from ${\mathbb F}_2$. This
matrix is called the generator matrix for the code ${\mathcal C}$.  The
generator matrix defines the encoding procedure for the code.  In particular if
we left multiply the $k \times n$ matrix $G$ by a $k$-tuple row vector of bits
to be encoded, then we obtain the encoded length $n$ string for these bits. For
a code ${\mathcal C}$, the parity check matrix, $P$ is an $(n-k) \times n$
matrix with elements from ${\mathbb F}_2$ such that $P v=0$ for all $v \in
{\mathcal C}$.  Note that we use a slightly non-conventional row and column
ordering for the parity and generator matrices.  The (Hamming) distance between
two elements of ${\mathbb F}_2$, $v$ and $w$, is given by $H(v,w)= \sum_{i=1}^n
\delta_{v_i,w_i}$, i.e. the number of places where the two $n$ bit strings
disagree.  We say that a classical linear code is a $[n,k,d]$ code if it
encodes $k$ bits into length $n$ bits and the minimum distance between any two
elements of the code space is $d$.

We now discuss quantum linear codes which are also known as stabilizer codes.
We will assume that the reader is familiar with stabilizer codes at the level
of \cite{Nielsen:00a} and provide the following review to set our definitions.

Suppose we have $n$ qubits. Then ${\mathcal P}$ is the group, commonly known as
the Pauli group, whose elements are made up of tensor products of single qubit
Pauli operators ($I,X,Y,Z$) along with a global phase of $i^s$, for $s \in
\{0,1,2,3\}$.  Every Pauli operator can be written in the form $i^s Z^{a_1}
X^{b_1} \otimes Z^{a_2} X^{b_2} \otimes \cdots \otimes Z^{a_n} X^{b_n}$ where
$s\in \{0,1,2,3\}$ and $a_i,b_i \in \{0,1\}$. The weight of a Pauli group
element is the number of non-identity single qubit operators in the above
expansion.  All Pauli group elements either commute with each other
$[P,Q]=PQ-QP=0$ or anticommute with each other $\{P,Q\}=PQ+QP=0$.

A stabilizer group is an abelian subgroup of the Pauli group which does not
contain the Pauli group element $-I^{\otimes n}$.  All of the elements of a
stabilizer square to $I^{\otimes n}$ and hence have eigenvalues $\pm 1$.  Given
a stabilizer group ${\mathcal S}$, we can define a stabilizer code as the
subspace of the Hilbert space on $n$ qubits which is stabilized by all of
elements in ${\mathcal S}$, i.e. the subspace ${\mathcal H}_S$ defined as all
states $|\psi\rangle$ such that $S|\psi\rangle=|\psi\rangle$ for all $S \in
{\mathcal S}$.  If the stabilizer group has a minimal set of generators
$S_1,\dots,S_{n-k}$, then the dimension of the stabilized subspace (the code
subspace) is $2^k$, i.e. it encodes $k$ qubits of quantum information.  The
normalizer ${\mathcal N}$ of the stabilizer ${\mathcal S}$ in ${\mathcal P}$ is
the set of elements of $P \in {\mathcal P}$ such that $P S P^\dagger \in
{\mathcal S}$ for all $S \in {\mathcal S}$.  Note that ${\mathcal S} \subseteq
{\mathcal N}$.  Elements of the normalizer preserve the stabilizer subspace,
$S_i (N |\psi\rangle)=N S_j |\psi\rangle=N |\psi\rangle$ for $N \in {\mathcal
N}$.  The group ${\mathcal N} / {\mathcal S}$ is a Pauli group on $k$ qubits,
i.e. ${\mathcal N}/{\mathcal S}$ acts as encoded operators on the stabilizer
code.

If an element $E$ of the Pauli group anticommutes with at least one stabilizer
element $S_k$ ($\{E,S_k\}=0$) then this element takes a state encoded into the
stabilizer subspace into one which is outside of this subspace.  This can be
verified by noting that for $|\psi\rangle \in {\mathcal H}_S$,
$E|\psi\rangle=ES_k |\psi\rangle=-S_k E |\psi\rangle$ which tells us that $E$
flips the sign of the eigenvalue of $S_k$.  This implies that if we encode
quantum information into our stabilizer subspace and then a Pauli operator $E$
which anticommutes with at least one $S_k$ acts on the system, then we can
detect this event by measuring the eignenvalue of $S_k$. Elements of the Pauli
group which anticommute with at least a single $S_k$ are elements of the Pauli
group which are not in ${\mathcal N}$.

This leads to the following characterization of the capabilities of a
stabilizer code to correct errors.  Let $\{E_i\}$ be a set of Pauli elements
such that $E_i^\dagger E_j \notin {\mathcal N} - {\mathcal S}$ for all $i,j$.
Then $\{E_i\}$ is a set of correctable errors for this stabilizer code.  The
weight of the smallest $E_i^\dagger E_j$ which is in ${\mathcal N}-{\mathcal
S}$ is the distance $d$ of the stabilizer code.  A quantum code which can
correct up to $t$ errors must have a distance at least $2t+1$.  A quantum error
correcting code which uses $n$ qubits, encodes $k$ qubits, and has a distance
$d$ is called a $[[n,k,d]]$ quantum code.

\section{Generalized Shor Codes}

Here we review codes which are simple generalizations of Shor's original
$[[9,1,3]]$ quantum error correcting code.  The codes we construct later in
this paper will be subsystem versions of these codes, achieving the same error
correcting properties as a generalized Shor code, but with significantly
reduction in the number of stabilizers which need to be measured in order to
perform quantum error correction.

To construct generalized Shor codes we first consider what a classical linear
code looks like in the stabilizer formalism.  Let ${\mathcal C}$ be a classical
linear code with generator matrix $G$ and parity check matrix $P$.  Suppose
that we wish to construct a quantum error correcting code which corrects bit
flips ($X$ errors) just as this classical code corrects bit flips. The
stabilizer for this code is then obtained simply from the rows of the parity
check matrix. In particular we can construct the stabilizer generators
\begin{equation}
S_i=\bigotimes_{j=1}^n Z^{P_{i,j}}.
\end{equation}
Clearly these stabilizer elements all commute with each other because they are
all made up of either $I$ or $Z$ Pauli operators, and they are independent of
each other since the rows of $P_{i,j}$ are linearly independent.  It is clear
that measuring these stabilizer generators is equivalent to using the parity
check matrix to calculate the syndrome for this stabilizer code.

What are the logical operators for this code?  The logical $X$ operators for
the $k$ encoded qubits can be obtained directly from the generator matrix, $G$.
In particular we can choose the $i$th logical $X$ operator to be
\begin{equation}
\bar{X}_i=\bigotimes_{j=1}^n X^{G_{i,j}}.
\end{equation}
These operators are certainly not in the stabilizer since they are made up
entirely of $X$ operators and they commute with all of the elements of the
stabilizer because
\begin{eqnarray}
\bar{X}_i S_j &=&\bigotimes_{k=1}^n X^{G_{i,k}} \bigotimes_{l=1}^n Z^{P_{j,l}}
\nonumber \\
&=&(-1)^{\sum_{k=1}^n G_{i,k} P_{j,k}} S_j \bar{X}_i=S_j\bar{X}_i
\end{eqnarray}
where we have used the fact that $\sum_{k=1}^n G_{i,k} P_{j,k}=0$.  Further we
see, as expected, there are $k$ of these encoded $\bar{X}$ operators.  What
about the encoded $\bar{Z}$ operators?  Well it is always possible to construct
these operators from tensor products of $Z$ and $I$ operators.  In particular
we let the $i$th logical $Z$ operator be
\begin{equation}
\bar{Z}_i=\bigotimes_{j=1}^n Z^{P_{i,j}^c}.
\end{equation}
What is $P_{i,j}^c$?  Let ${\mathcal S}$ be the subspace of $({\mathbb F}_2)^n$
spanned by the rows of the parity check matrix $P$.  Then $P^c$ is a $k$ by $n$
matrix whose row vectors are linearly independent from the rows of $P$, are
themselves linearly independent, and which together with the rows of $P$ span
the entire space $({\mathbb F}_2)^n$.  In other words $P^c$ is made up of rows
which together with the rows of $P$ form a basis for the full space $({\mathbb
F}_2)^n$.  The choice of a generator matrix $G$ forces a particular choice (up
to row multiplications) for $P^c$ such that it is a logical $\bar{Z}_i$
operator.  Encoded $Y$ operators can be obtained directly from
$\bar{Y}_i=i\bar{X}_i \bar{Z}_i$.

In a similar vein to the construction of $P^c$, we can define a $n-k$ by $n$
matrix $G^c$ such that the rows of this matrix are linearly independent of the
other rows of $G$, linearly independent of each other, and together with the
rows of $G$ form a basis for $({\mathbb F}_2)^n$.  We can then define the
operators
\begin{equation}
E_i=\bigotimes_{j=1}^n X^{G^c_{i,j}}.
\end{equation}
Now the set of detectable bit flip errors is easily defined.  It is any error
which can be expressed as a product of at least one $E_i$ and any number
(including zero) of $\bar{X}_i$ operators.  The reason these are detectable is
that products of $E_i$ operators are guaranteed to anticommute with at least
one $S_i$.

Table \ref{table:code} below  presents a useful way of thinking about the
structure of classical linear codes in the stabilizer formalism.    In this
table, each box shows the matrices used to construct a group for the linear
classical code and the number of independent generators of the group is listed
at the top of each column. In addition to the stabilizer and encoded $X$ and
$Z$ operators, another group is formed from $G^c$.  We have labelled this
latter group the group of ``pure errors.'' These are errors which do not affect
the encoded information and are detectable errors.

\begin{table}[h]
\caption{Classical linear codes in the stabilizer formalism.}\label{table:code}
\begin{tabular}{l|l|l|}

 & $n-k$ Generators & $k$ Generators  \\
\cline{1-3}
  $\begin{array}{l}
  {\rm Tensor~product~of~}\\
  Z {\rm~and~} I{\rm ~operators}
   \end{array}$
& $P$ (Stabilizer) & $P^c$ (Encoded $Z$) \\
  \cline{1-3}
  $\begin{array}{l}
  {\rm Tensor~product~of~}\\
  X {\rm~and~} I{\rm ~operators}
   \end{array}$
 & $G^c$ (Pure Errors) & $G$ (Encoded $X$) \\
\cline{1-3}
\end{tabular}
\\
\\
\end{table}

Finally let us note that the above construction can be used to correct phase
flip errors instead of bit flip errors.  This can be done by interchanging the
$Z$ Pauli operators in the stabilizer with the $X$ Pauli operators.  Replacing
$Z$ Pauli operators in the logical $Z$ operator with the $X$ Pauli operators
will produce a new encoded $Z$ operator for this code.  Replacing $X$ Pauli
operators in the logical $X$ operator with the $Z$ Pauli operators will produce
a new encoded $X$ operator for this code.

We are now ready to describe generalized Shor codes.  Shor's basic idea was
that one could construct a quantum error correcting code for errors on qubits
by concatenating a code designed to deal with bit flips ($X$ errors) with one
which is designed to deal with phase flips ($Z$ errors).  Notice that there is
an asymmetry in this construction: one should decide which of these errors to
be dealt with on the lowest level of the concatenation and which should be
dealt with at the second level of the concatenation.

Let ${\mathcal C}_1$ and ${\mathcal C}_2$ be two $[n_1,k_1,d_1]$ and
$[n_2,k_2,d_2]$ linear codes. Let ${\mathcal C}_1$ and ${\mathcal C}_2$ have
generator matrices, $G_1$ and $G_2$ respectively, and have parity check
matrices $P_1$ and $P_2$ respectively.  In generalized Shor codes we use these
two codes to construct a $[n_1 n_2, k_1 k_2, {\rm min}(d_1,d_2)]$ quantum error
correcting code.  To do this we proceed as follows.  Take $n_1 n_2$ qubits and
partition them into $n_2$ blocks of size $n_1$.  For each of these blocks of
$n_1$ qubits we can define a quantum error correcting code from ${\mathcal
C}_1$ which is designed to correct bit flip errors as above.  Each of the $n_1$
blocks encodes $k_1$ qubits.  We can now use these encoded qubits to protect
against phase flip errors. The first observation we need is than any tensor
product of $Z$ and $I$ acting as an error on one of our blocks will act as a
phase error on one or more of the encoded qubits times an element of the
stabilizer. The reason for this is that every tensor product of $Z$ and $I$ is
either in the stabilizer of quantum error correcting or is a product of a
stabilizer operator and encoded $Z$ operators.  This allows us to consider
phase errors on blocks as phase errors on the encoded qubit.  To protect
against phase flip errors we begin by picking the $i$th encoded qubit from
every block. We can then use each of these from all blocks to construct a
quantum error correcting code for phase flips using the ${\mathcal C}_2$ code.
Since for each choice of $i$ we obtain a code with $k_2$ encoded qubits, this
code can be used to store $k_1k_2$ qubits. Further, the code will be of
distance $d_1$ for bit flip errors and $d_2$ for phase flip errors.  For $Y$
errors the code will be of distance ${\rm min}(d_1,d_2)$. Thus the distance of
the code will be ${\rm min}(d_1,d_2)$.

Generalized Shor codes are easy to construct, but do not have nice asymptotic
error correcting properties.  However, they are conceptually extremely easy to
understand and their error recovery routines are easily obtained from their
constituent classical linear code recovery routines. One important point about
these codes, however, is that they have an asymmetry in their recovery routines
due to the choice of whether to concatenate bit flip codes with phase flip
codes or vice versa.  In particular for each the lowest level of concatenation,
error correction should be performed for every single block. Indeed it is easy
to see that the stabilizer for the generalized Shor code constructed above is
generated by a set of $(n_1-k_1)n_2+(n_2-k_2)$ independent operators. We will
show below that using the notion of a subsystem, this can be reduced, for every
generalized Shor code, to $(n_1-k_1)k_2+(n_2-k_2)k_1 $.

\section{Quantum Error Correcting Subsystem Codes}

In stabilizer codes one encodes information into a subspace of a quantum
system.  However the most general way to encode information is not to encode it
into a subspace, but instead to encode it into a subsystem\cite{Knill:00a}.
Let us briefly review this concept and describe the notion of quantum error
correcting subsystem codes.

Suppose we have a Hilbert space ${\mathcal H}$.  Then one method for encoding
quantum information is to encode this information into a subspace of ${\mathcal
H}$.  In particular if ${\mathcal H}$ is the direct sum of two subspaces,
${\mathcal H}_C$ and ${\mathcal H}_D$, ${\mathcal H}={\mathcal H}_C \oplus
{\mathcal H}_D$, then we can encode quantum information into one of the
subspaces, ${\mathcal H}_C$.   In addition to the notion of a direct sum,
$\oplus$, of two Hilbert spaces, another notion for combining two Hilbert
spaces is to construct the tensor product of these two Hilbert spaces.  Thus,
for example, we can combine two Hilbert space ${\mathcal H}_c$ and ${\mathcal
H}_D$ as ${\mathcal H}={\mathcal H}_C \otimes {\mathcal H}_D$.  Then we can
encode quantum information into the {\em subsystem} ${\mathcal H}_C$.  Notice
that such an encoding, for a fixed encoding into ${\mathcal H}_D$ is a subspace
encoding, but, without such a specification, the encoding is not a subspace
encoding.

By repeatedly constructing subsystems and subspaces on Hilbert spaces, we can,
most generally decompose a Hilbert space into a multiple direct sum of multiple
tensor products of Hilbert spaces (since the process of direct sum and tensor
product obey a distributive law.) Further if we single out a single one of
these Hilbert spaces, call it ${\mathcal H}_C$, then we may collect the other
terms in such a decomposition so that the Hilbert space decomposes as
\begin{equation}
{\mathcal H}=({\mathcal H}_C \otimes {\mathcal H}_D) \oplus {\mathcal H}_E.
\end{equation}
The decomposition described above, which is the most general for encoding a
single Hilbert space, can be described as taking a Hilbert space ${\mathcal H}$
and decomposed it into a subspace ${\mathcal H_E}$ and a perpendicular
subspace, ${\mathcal H}_E^\perp$. On this perpendicular subspace we have
further decomposed this into a tensor product of two subsystem Hilbert spaces,
${\mathcal H}_E^\perp={\mathcal H}_C \otimes {\mathcal H}_D$.  Thus if we are
going to encode quantum information to the subsystem ${\mathcal H}_C$ we can do
this by preparing the quantum state
\begin{equation}
\rho=(\rho_C \otimes \rho_D) \oplus 0_E
\end{equation}
where $\rho_C$ is the density matrix of the encoded quantum information,
$\rho_D$ is information encoded into the subsystem ${\mathcal D}$ (which can be
arbitrary) and $0_E$ is the all zero matrix on the subspace ${\mathcal H}_E$.
At this point we can see one of the particular features of subsystem encodings:
if we act nontrivially on the subsystem ${\mathcal D}$ then the quantum
information encoded into the subsystem ${\mathcal C}$ is not affected. In other
words, information encoded into a subsystem is not affected by information
encoded into different subsystems.  Encoding into subsystems has been used most
notably in noiseless subsystems\cite{Knill:00a,Kempe:01a,Zanardi:01a} and
communicating without a shared reference frame\cite{Bartlett:03a}, as well as
being essential to an important transform in quantum information theory, the
quantum Schur transform\cite{Bacon:04a,Bacon:06b}.

What does this mean for the theory of quantum error correction?  Suppose that
we encode quantum information into a subsystem ${\mathcal H}_C$ of a Hilbert
space which has been decomposed as ${\mathcal H}=({\mathcal H}_C \otimes
{\mathcal H}_D ) \oplus {\mathcal H}_E$.  Next suppose that a quantum
operation, corresponding to some quantum error, occurs on our system.  Then the
goal of quantum error correction is to restore the information encoded into the
subsystem ${\mathcal H}_C$.  In the case where we have a subspace code, i.e.
when ${\mathcal H}_D=0$, then we must apply an operation which correctly
restores the quantum information encoded into the subspace ${\mathcal H}_C$. If
on the other hand we have a subsystem code, ${\mathcal H}_D \neq 0$, then we
must apply an operation which correctly restores the information encoded into
the subsystem ${\mathcal H}_C$, but we do not care what happens in this
procedure to the information encoded into ${\mathcal H}_D$.  In other words, if
we are to perform quantum error correction on a subsystem code, then the error
recovery routine need only correct the error modulo the subsystem structure. We
need not be worried if information encoded into ${\mathcal H}_D$ is destroyed
by the entire error/recovery routine, as long as the information in ${\mathcal
H}_C$ is correctly restored.

Suppose that we wish to protect our quantum information from a set of errors
${E_a}$ after we have encoded the quantum information into a subsystem as
described above.   Suppose that $|i\rangle \otimes |j\rangle$ is a basis for
the subspace ${\mathcal H}_C \otimes {\mathcal H}_D$, then a necessary and
sufficient condition\cite{Kribs:05a,Nielsen:05a} for the set of errors $E_a$ to
be correctible is that
\begin{equation}
(\langle i| \otimes \langle k| ) E_a^\dagger E_b (|j\rangle \otimes |l\rangle)
= \delta_{i,j} c_{a,b}.
\end{equation}
Notice that this condition does not depend on the $|k\rangle$ and $|l \rangle$.

For a more complete description of quantum error correcting subsystems we refer
the reader to \cite{Kribs:05a,Kribs:05b} which details not just the notion of a
quantum error correcting subsystem, but also the notion of a operator quantum
error correction, which is a complete method for dealing with quantum error
correcting subsystems.

\section{Subsystem Codes from Two Linear Codes}

We now turn to the construction of a new class of quantum error correcting
subsystems.  We will begin by detailing the construction and then proving that
our construction has the error correcting properties which we claim.  Our
construction follows, in rough outline, that presented in \cite{Bacon:06a}.

\subsection{Subsystem Code Construction}

Suppose we are given two classical linear codes, ${\mathcal C}_1$ and
${\mathcal C}_2$, which are a $[n_1,k_1,d_1]$ code and a $[n_2,k_2,d_2]$ code,
respectively.  We will now show how to use these codes to construct a quantum
error correcting subsystem code which is a $[[n_1n_2, k_1 k_2, {\rm
min}(d_1,d_2)]]$ code.  Let $P_1$ and $P_2$ denote the parity check matrices
and $G_1$ and $G_2$ denote the generator matrices for the two codes ${\mathcal
C}_1$ and ${\mathcal C}_2$ respectively.  From the rows of $P_1$ we can
construct a stabilizer code on $n_1$ qubits. In particular we can define the
$n_1-k_1$ stabilizer operators $S_i = \otimes_{j=1}^{n_1} Z^{(P_1)_{ij}}$. Call
the stabilizer group generated by this set of operators ${\mathcal S}_1=
\langle S_1,\dots,S_{n_1-k_1} \rangle$. In a similar manner we can use $P_2$ to
construct a stabilizer code. In particular define the $n_2-k_2$ stabilizer
operators $T_i=\otimes_{j=1}^{n_2} X^{(P_2)_{ij}}$.  This forms a stabilizer
group ${\mathcal S}_2 = \langle T_1,\dots, T_{n_2-k_2} \rangle$.

These codes are classical codes except that the second code is not in the
computational basis $|0\rangle$, $|1\rangle$, but is instead in the dual
$|+\rangle$, $|-\rangle$ basis.  The first code is designed to correct $\lfloor
{d_1-1 \over 2} \rfloor$ bit flip errors (Pauli $X$ errors) while the second
code is designed to correct $\lfloor {d_2-1 \over 2} \rfloor$ phase flip errors
(Pauli $Z$ errors.)  For both codes, we can follow our construction of
classical codes and construct encoded operators.  For the first code call the
encoded $X$ and $Z$ operators $(\bar{X}_1)_i$ and $(\bar{Z}_1)_i$ and for the
second code call the encoded $X$ and $Z$ operators $(\bar{X}_2)_i$ and
$(\bar{Z}_2)_i$.

Now we will show how to produce a subsystem code using these codes.  Put $n=n_1
n_2$ qubits on a rectangular $n_1 \times n_2$ lattice (this lattice is for
illustrative purposes only, and is not a necessary part of the code.) We will
now use the stabilizer codes operators ${\mathcal S}_1$ in the columns and the
stabilizer code operators ${\mathcal S}_2$ in the rows to construct a
nonabelian group ${\mathcal T}$ on these $n^2$ qubits. More specifically, let
${\mathcal T}_1$ be the stabilizer group made up of letting ${\mathcal S}_1$
operators acting on all of the columns of the lattice and let ${\mathcal T}_2$
be the stabilizer group made up of letting ${\mathcal S}_2$ operators acting on
all of the rows of the lattice. Then the group we are considering, ${\mathcal
T}$, is the group generated by the elements of ${\mathcal T}_1$ and ${\mathcal
T}_2$.

The group ${\mathcal T}$ is clearly nonabelian.  From ${\mathcal T}$ we can
further construct an abelian invariant subgroup (invariant meaning that all
elements of ${\mathcal T}$ commute with the elements of the subgroup.)  To do
this, we do the following.  Take one of the stabilizer operators from
${\mathcal S}_1$. Take one of the codewords from ${\mathcal C}_2$, call it $v$.
Now construct an operator on our $n$ qubits which has $S_1^{v_j}$ acting on
each column $j$ (where $S_1^0=I$.)  Clearly these elements are in ${\mathcal
T}$. Further they commute with all of the elements of ${\mathcal T}$ since in a
particular {\em row} they are made up of $Z$ operators which have the form
$\otimes_{j=1}^{n_2} Z^{v_j}$ and in a particular column they are made up of
elements of ${\mathcal S}_1$.  We can similarly take stabilizer operators form
${\mathcal S}_2$ and codewords $w$ from ${\mathcal C}_1$ and construct
operators which act like $S_2^{w_j}$ acting on each row $j$.  These operators
will also commute with all of the elements of ${\mathcal T}$.  So now we can
form our abelian invariant subgroup ${\mathcal S}$ as the group generated by
these stabilizer operators for all possible stabilizer codeword combinations in
both rows and columns.

So now we have a structure set up where we have a non-abelian group ${\mathcal
T}$ and an abelian invariant subgroup of this group, ${\mathcal S}$.  There is
another set of operators which are important, which will correspond to the
logical operators on the code ${\mathcal L}$.  Suppose we take an encoded $X$
operator for the stabilizer code ${\mathcal S}_1$, call it $(\bar{X}_1)_i$, and
take an encoded operator for the stabilizer ${\mathcal S}_2$, call it
$(\bar{X}_2)_j$.  Then we can form an operator $\bar{X}_{i,j}$ acting on our
$n^2$ qubits by putting $(\bar{X}_1)_i$ in each column $j$ where
$(\bar{X}_2)_j$ acts non-trivially as $X$.  Similarly from $(\bar{Z}_1)_i$ and
$(\bar{Z}_2)_j$ we can construct an operator $(\bar{Z}_{i,j})$ which is
$(\bar{Z}_1)_i$ in each column $j$ where $(\bar{Z}_2)_j$ acts non-trivially as
$Z$.  It is easy to see that $\bar{X}_{i,j}$ and $\bar{Z}_{i,j}$ commute with
the group ${\mathcal T}$ since in each row or column they look like encoded
operators.  Further these operators also anticommute with each other if and
only if their indices match,
$\{\bar{X}_{i,j},\bar{Z}_{k,l}\}=\delta_{i,k}\delta_{j,l}$, and hence commute
with each other otherwise.  These operators further form a group which is
isomorphic to a Pauli group on $k_1 k_2$ qubits.

Next we need to discuss how to put ${\mathcal T}$, ${\mathcal S}$ and
${\mathcal L}$ together to form a subsystem code.  This is nearly identical to
the procedure described in \cite{Bacon:06a}.

To do this, it is convenient to adopt explicit forms for the operators we have
described above in a simple notation.  Let $M$ be a $n$ by $n$ matrix with
entries either $0$ or $1$.  Then we define $P^{M}$ as the operator on our $n$
by $n$ qubits which is a tensor product of $P$ and $I$ operators which acts on
the qubit at the $i$th row and $j$th column as $P^{M_{i,j}}$.  In this notion
it is easy to see that every element of the Pauli group on our $n^2$ qubits can
be expressed as $i^k Z^{A}X^{B}$ where $k \in \{0,1,2,3\}$ and $A$ and $B$ are
$n$ by $n$ $0/1$ matrices.  We will now proceed to use this notation to express
the operators in ${\mathcal T}$, ${\mathcal S}$ and ${\mathcal L}$.

First consider ${\mathcal T}$.  Consider operators in ${\mathcal T}$ which have
elements of $S_1$ which lie in a column and are products of $Z$ operators. In
our new notation, these operators can be written as $Z^{A}$ where
$A_{i,j}=(p_1^T P_1)_i \delta_{j,j_0}$ where $p_1^T$ is a $n_1-k_1$ binary row
vector and $j_0$ is the column where this operator acts. Similarly, we can
construct operators from $S_2$ lie in a row and are products of $X$ operators.
They can be expressed as $X^B$ where $B_{i,j}=\delta_{i,i_0} (p_2^T P_2)_j$
where $i_0$ is the row where this operator acts and $p_2^T$ is a $n_2-k_2$
binary row vector. In order to make a distinction which will be useful later,
it is useful to express the delta functions in the above expression as follows.
Since the rows of $G_1$ and $G_1^c$ form a basis for the entire space
$({\mathbb F}_2)^n$ we can express $\delta_{j,j_0}$ as $(g_1^T G_1 + (g_1^c)^T
G_1^c)_j$ for some choice of length $k$ binary row vector $g_1^T$ and length
$n-k$ binary row vector $(g_1^c)^T$. We can perform a similar decomposition for
the $X$ operators.  Since multiplication now corresponds, up to a phase factor,
to addition, it is then easy to see that every element of the group ${\mathcal
T}$ can be expressed as $t(Q,Q^c,R,R^c,p)=i^p Z^{A} X^{B}$ with
\begin{equation}
A= P_1^T (Q) G_2 + P_1^T (Q^c) G_2^c
\end{equation}
and
\begin{equation}
B= G_1^T (R) P_2 + (G_1^c)^T (R^c) P_2
\end{equation}
where $Q$ is a $n_1-k_1$ by $k_2$ $0/1$ matrix, $Q^c$ is a $n_1-k_1$ by
$n_2-k_2$ $0/1$ matrix, $R$ is a $k_1$ by $n_2-k_2$ $0/1$ matrix, $R^c$ is a
$n_1-k_1$ by $n_2-k_2$ $0/1$ matrix, and $p \in \{0,1,2,3\}$.  Note that we
have extended the group slightly by adding in a phase factor of $i$.  This is
for convenience sake when describing certain Pauli subgroups of ${\mathcal T}$.

Having described ${\mathcal T}$ in our notation, we now turn to ${\mathcal S}$.
${\mathcal S}$ is a subgroup of ${\mathcal T}$ and thus we can express it again
as a $t(Q,Q^c,R,R_c,p)$.  In particular from the definition of the ${\mathcal
S}$ we see that its elements are of the form $s(Q,R)=t(Q,0,R,0,0)$, i.e. they
are elements of ${\mathcal T}$ with $Q^c=0$, $R^c=0$, and $p=0$.

Next we express elements of ${\mathcal L}$ in this notation.  Elements of
${\mathcal L}$ are formed from encoded operations from the two codes.  In
particular it is easy to see that they can be expressed as $l(U,V,p)=i^p Z^A
X^B$ with
\begin{equation}
A=(P_1^c)^T (U) G_2
\end{equation}
and
\begin{equation}
B=G_1^T (V) P_2^c
\end{equation}
where $U$ and $V$ are $k_1$ by $k_2$ $0/1$ matrices and $p\in \{0,1,2,3\}$.
Again we have added an extra phase factor of $i$.

Finally let us notice that every element of the Pauli group on our $n^2$ qubits
can be expressed as $o(Q,Q^c,R,R^c,U,U^c,V,V^c,p)=i^p Z^A X^B$ with
\begin{equation}
A=P_1^T (Q) G_2 + P_1^T (Q^c) G_2^c+(P_1^c)^T (U) G_2+(P_1^c)^T (U^c) G_2^c
\end{equation}
and
\begin{equation}
B= G_1^T (R) P_2 + (G_1^c)^T (R^c) P_2+G_1^T (V) P_2^c+(G_1^c)^T (V^c) P_2^c
\label{eq:Bexpansion}
\end{equation}
where $Q$, $Q^c$, $R$, $R^c$, $U$, and $V$ have the dimensions listed above,
$U^c$ is a $k_1$ by $n_2-k_2$ $0/1$ matrix, $V^c$ is a $n_1-k_1$ by $k_2$ $0/1$
matrix, and $p \in \{0,1,2,3\}$.

We can now define our subsystem code.  We begin with ${\mathcal S}$.
${\mathcal S}$ is an abelian subgroup of the Pauli group which does not contain
$-I^{\otimes n^2}$. Thus it is a stabilizer group.  From this stabilizer group
we can from a stabilizer code ${\mathcal S}$.  Operators which are in
${\mathcal T}$ and ${\mathcal L}$ are then in the normalizer of this stabilizer
code, since all of the elements of ${\mathcal T}$ and ${\mathcal L}$ commute
with all of the elements of ${\mathcal S}$.  Further we can think of the
operators from ${\mathcal L}$ and ${\mathcal T}$ acting on different logical
qubits for the stabilizer code, since all of the elements of ${\mathcal L}$
commute with all of the elements of ${\mathcal T}$.

The subsystem code can now be defined.  Since ${\mathcal S}$ is a stabilizer
code, we can label subspaces of the $n^2$ qubits Hilbert space by the $\pm 1$
eigenvalues of the stabilizer generators.  There are $(n_1-k_1)k_2$ such
stabilizer generators made up of tensor products of $I$ and $Z$ operators and
$k_1(n_2-k_2)$ such stabilizer generators made up of tensor products of $I$ and
$X$ operators.  Thus the number of generators for this stabilizer group is
$(n_1-k_1)k_2+k_1(n_2-k_2)$.  Next notice that ${\mathcal L}$ forms an encoded
Pauli group acting on $k_1 k_2$ qubits.  To see this notice that $l(U,v,p)$
follows the multiplication rules of a Pauli group as if for $k_1k_2$
independent qubits.  Now consider ${\mathcal T}/{\mathcal S}$, the group
${\mathcal T}$ after we divide out the stabilizer group.  This group is made up
of operators $t(0,Q^c,0,R^c,p)$.  This group is an encoded Pauli group on
$(n_1-k_1)(n_2-k_2)$ qubits.  Putting this together we have a stabilizer with
$(n_1-k_1)k_2+k_1(n_2-k_2)$ generators, encoded ${\mathcal T}/{\mathcal S}$
operators which act as a Pauli group on $(n_1-k_1)(n_2-k_2)$ qubits, and
encoded ${\mathcal L}$ operators which act as a Pauli group on $k_1 k_2$
qubits.  The total of these generators and the number of encoded qubits is
$(n_1-k_1)(n_2-k_2)+(n_1-k_1)k_2+k_1(n_2-k_2)+k_1 k_2=n_1 n_2$.  This implies
that the ${\mathcal T}/{\mathcal S}$ encoded operators and the ${\mathcal L}$
encoded operators form an exhaustive list of encoded operators for the
stabilizer code ${\mathcal S}$.

So how do we define our subsystem code?  Since there are
$(n_1-k_1)k_2+k_1(n_2-k_2)$ stabilizer generators, we can label subspaces of
dimension $2^{n_1 n_2-(n_1-k_1)k_2+k_1(n_2-k_2)}$ by the $\pm 1$ eigenvalues,
call them $s_i$, of these stabilizer generators.  Call $s$ the
$(n_1-k_1)k_2+k_1(n_2-k_2)$ tuple of these values.  Further for each such
subspace there is now a tensor product between encoded logical operators from
${\mathcal L}$ and those from ${\mathcal T}/{\mathcal S}$.  Thus we can find a
basis for our Hilbert space on $n_1 n_2$ qubits such that it decomposes as
\begin{equation}
{\mathcal H}=\bigoplus_{s \in \{0,1\}^{(n_1-k_1)k_2+k_1(n_2-k_2)}} {\mathcal
H}_s^{\mathcal T/S} \otimes {\mathcal H}_s^{\mathcal L}
\end{equation}
where ${\rm dim} {\mathcal H}_s^{\mathcal L}=2^{k_1 k_2}$ and ${\rm dim}
{\mathcal H}_s^{\mathcal T}=2^{(n_1-k_1)(n_2-k_2)}$.  It is now useful to
describe how elements of ${\mathcal S}$, ${\mathcal T/S}$ and ${\mathcal L}$
operate on this decomposition.  Elements of ${\mathcal S}$ act as either $\pm
1$ on each subspace.  In particular they act as
\begin{equation}
\bigoplus_{s \in \{0,1\}^{(n_1-k_1)k_2+k_1(n_2-k_2)}}
(-1)^{\sum_{i=1}^{(n_1-k_1)k_2+k_1(n_2-k_2)} s_i} I \otimes I
\end{equation}
Elements of ${\mathcal L}$ act as encoded qubits on the ${\mathcal
H}_s^{\mathcal L}$ subsystems
\begin{equation}
\bigoplus_{s \in \{0,1\}^{(n_1-k_1)k_2+k_1(n_2-k_2)}}
 I \otimes L(s)
\end{equation}
while elements of ${\mathcal T}$ act as encoded qubits on the ${\mathcal
H}_s^{\mathcal T}$ subsystems
\begin{equation}
\bigoplus_{s \in \{0,1\}^{(n_1-k_1)k_2+k_1(n_2-k_2)}}
 T(s) \otimes I.
\end{equation}
Our subsystem code can now be defined.  We will encode our quantum information
into the all $s_i=+1$ subspace and the corresponding ${\mathcal H}_s^{\mathcal
L}$ subsystem.  This encoding will encoded $k_1 k_2$ qubits and the logical
Pauli operators on this code come from ${\mathcal L}$.  Note that elements of
${\mathcal T}$ can always be expressed as operators which do not act on this
subsystem.  This is a subsystem degree of freedom which makes our code a
quantum error correcting subsystem code.

\subsection{Subsystem Error Correcting Routine}

Having identified the subsystem we are encoding into and the representation
theoretic structure of operators in ${\mathcal S}$, ${\mathcal T}$ and
${\mathcal L}$ we can now turn to the error correcting procedure for this code.
We will show that this code can detect single qubit errors of weight $\min
(d_1,d_2)$ and describe the error recovery routine for this code.  This error
recovery routine corrects the information in our subsystem but may act
nontrivially on the subsystem ${\mathcal H}_s^{\mathcal T}$.

Suppose that ${\mathcal C}_1$ and ${\mathcal C}_2$ can correct the sets of
errors ${\mathcal E}_1$ and ${\mathcal E}_2$ respectively.  We will now show
how this allows us to correct bit flip and phase flip errors modulo the
subsystem structure.

First consider an $X$ error on our code.  Suppose that $e_1$ is a correctable
error for code ${\mathcal C}_1$.  Let $E_1$ be a $n_1$ by $n_2$ matrix.  We
will show that every error $X^B$ with $B_{i,j}=(e_1)_i (E_1)_{i,j}$ is a
correctable error for our subsystem code.  To see this we first note that if we
express this error as in Eq.~(\ref{eq:Bexpansion}) as
\begin{equation}
B= G_1^T (R) P_2 + (G_1^c)^T (R^c) P_2+G_1^T (V) P_2^c+(G_1^c)^T (V^c) P_2^c
\end{equation}
then we can turn this into a product of an element of ${\mathcal T}$ and one
which is not in ${\mathcal T}$, $X^{B_1} X^{B_2}$ where
\begin{equation}
B_1= G_1^T (V) P_2^c+(G_1^c)^T (V^c) P_2^c,
\end{equation} and
\begin{equation}
B_2= G_1^T (R) P_2 + (G_1^c)^T (R^c) P_2.
\end{equation}
Since $X^{B_2}$ is an element of ${\mathcal T}$ it cannot act as an error on
the information encoded into our subsystem.  Therefore we can consider the
error to be purely of the form $X^{B_1}$.  Now since the errors we are
considering have $B$ matrices of the form $X^B$ with $B_{i,j}=(e_1)_i
(E_1)_{i,j}$ this means that we can restrict the rows of $E_1$ to be from the
subspace spanned by $P_2^c$.

Now suppose that we measure the elements of ${\mathcal S}$ made up of tensor
product of $I$ and $Z$ operators.  These operators are of the form $Z^A$ where
\begin{equation}
A=(P_1)^T (U) G_2
\end{equation}
$P_2^c$ and $G_2$ act as encoded $X$ and $Z$ operators, respectively, for the
quantum version of the code ${\mathcal C}_2$. Thus by measuring the stabilizer
generators which are tensor products of $I$ and $Z$ operators, we can, for each
encoded qubit in the second code, make a measurement of the $P_1$ for these
encoded qubits.  If the error $e_1$ is a correctable error for $C_1$, then we
can apply the appropriate $e_1 p_i^T$ operators where $p_i$ are the appropriate
row vectors from $P_2$.  The effect of this correction procedure will be to
restore the information encoded into the subsystem up to the operator $X^{B_2}$
which is an element of ${\mathcal T}$.  Thus we see that, as claimed, that we
can correct errors which are of the form $X^B$ with $B_{i,j}=(e_1)_i
(E_1)_{i,j}$.

A similar prescription applies for $Z$ errors by measuring the stabilizer
generators which are made up of tensor products of $I$ and $X$ operators.  If
$e_2$ is a correctable error for the code ${\mathcal C}_2$, then this will
correct errors of the form $Z^A$ with $A_{i,j}=(e_2)_j (E_2)_{i,j}$.  If these
two procedures are carried out one after another they will also correct errors
which $Z^A X^B$ with $A$ and $B$ correctable as above.

Thus we see that by measuring the generators of ${\mathcal S}$ we can correct
errors related to the original codes ${\mathcal C}_1$ and ${\mathcal C}_2$.
What is the distance of this code?  For bit flip errors the distance will be
$d_1$ since we the smallest error of the form $X^B$ with $B_{i,j}=(e_1)_i
(E_1)_{i,j}$ has only single $X$ errors in a row.  Similarly the distance for
phase flip errors will be $d_2$.  The full distance must include $Z^A X^B$
errors and thus the distance is $\min (d_1,d_2)$.

\subsection{Savings Over Generalized Shor Codes}

In a generalized $[[n_1n_2, k_1 k_2, \min(d_1,d_2)]]$ Shor code we have seen
that error correction is achieved by measuring $(n_1-k_1)n_2+(n_2-k_2)$
stabilizer generators.  In our construction of a $[[n_1n_2, k_1 k_2,
\min(d_1,d_2)]]$ subsystem code above we have achieved the same parameters for
the code but now using only $(n_1-k_1)k_2+k_1(n_2-k_2)$ stabilizer
measurements.  Both methods use two classical linear error correcting codes to
construct a new quantum error correcting code.  Indeed, the subsystem codes we
present are nothing more than generalized Shor codes with certain stabilizers
which do not add to the error correcting distance removed\cite{Poulin:05a}. The
subsystem code versions of generalized Shor codes have considerable advantages
over the subspace code when it comes to the complexity of the error recovery
routine, providing a quadratic savings in the number of stabilizers which need
to be measured.

\section{Examples}

In this section we present a few examples of our code construction.

\subsection{Redundancy Code}

This is the construction presented in \cite{Bacon:06a}.  Let ${\mathcal C}_1$
and ${\mathcal C}_2$ both be a simple $n$ qubit redundancy code with generator
matrix $G=(1,1,\dots,1)$ and parity check matrix
\begin{equation}
P=\left[ \begin{array}{ccccc} 1 & 1 & 0 & \cdots & 0 \\
0 & 1 & 1 & \cdots & 0 \\
\vdots & &\ddots & & \vdots \\
0 & \cdots & 0 & 1 &1\end{array} \right]
\end{equation}
This redundancy code is a $[n,1,n]$ code.  The resulting subsystem code is a
$[[n^2,1,n]]$ code.  Properties of this code are described further in
\cite{Bacon:06a}.  Notice that a generalized Shor code constructed from this
redundancy code requires the measurement of $n^2-1$ stabilizer operators.  Our
subsystem code achieves the same parameters for the code but using measurements
of only $2(n-1)$ stabilizer operators.  This increase in efficiency combined
with other nice properties of this code has recently been shown to improve the
threshold for fault-tolerant quantum computation\cite{Aliferis:06a}.

\subsection{A [[49,1,5]] Code Which Outperforms a Concatenated Steane Code}

Consider using the Hamming $[7,4,3]$ code for the codes ${\mathcal C}_1$ and
${\mathcal C}_2$.  Our construction will yield a quantum error correcting
subsystem code with parameters $[[49,16,3]]$.  Now consider using those $16$
encoded qubits in a redundancy $[[16,1,4]]$ code as described in the last
subsection. If we use the error recovery for our $[[49,16,3]]$ code followed by
error correction for the $[[16,1,4]]$ code this will produce a code which can
correct arbitrary $2$ qubit errors and is thus effectively a $[[49,1,5]]$ code.
The number of stabilizers which need to be measured for this use of these codes
is $24+6=30$.  Another option is to use on $9$ of the $16$ encoded operators in
the redundancy subsystem code described above (allowing any error to occur on
the other subsystem.)  In this case one achieves a subsystem code with
parameters $[[49,1,5]]$ but with a number of stabilizers given by $24+4=28$.

This should be compared with the normal (not optimal) use of concatenating the
Steane code $[[7,1,3]]$ code\cite{Steane:98a} with itself.  There one uses
error correction on the Steane code $[[7,1,3]]$ for multiple levels of the
concatenation.  This results in a code which is effectively a $[[49,1,5]]$
code.  This concatenation scheme will require the measurement of measure
$42+6=48$ stabilizer operators. Thus we see that a considerable savings of $18$
(or $20$) less stabilizer measurements.

\section{Conclusion}

We have shown that for every generalized Shor code there is an subsystem code
with the same parameters but which requires significantly fewer stabilizer
measurements in order to perform quantum error correction.  These codes are
generalization of the codes presented in \cite{Bacon:06a} and are in the class
of stabilizer subsystem codes described in \cite{Poulin:05a}.  Recently
Aliferis and Cross \cite{Aliferis:06a} have used the subsystem codes described
in \cite{Bacon:06a} to significantly improve the provable threshold for
fault-tolerant quantum computation.  A major open question is whether the
subsystem codes described in the work described here can lead to similar and
perhaps greater increases in the threshold for fault-tolerant quantum
computation.

\section{Acknowledgements}

DB acknowledges funding from NSF grant number 0523359 and NSF grant number
0621621.


\bibliographystyle{IEEEtran}
\bibliography{IEEEabrv,dimdoub}

\begin{thebibliography}{10}
\providecommand{\url}[1]{#1}
\csname url@rmstyle\endcsname
\providecommand{\newblock}{\relax}
\providecommand{\bibinfo}[2]{#2}
\providecommand\BIBentrySTDinterwordspacing{\spaceskip=0pt\relax}
\providecommand\BIBentryALTinterwordstretchfactor{4}
\providecommand\BIBentryALTinterwordspacing{\spaceskip=\fontdimen2\font plus
\BIBentryALTinterwordstretchfactor\fontdimen3\font minus
  \fontdimen4\font\relax}
\providecommand\BIBforeignlanguage[2]{{%
\expandafter\ifx\csname l@#1\endcsname\relax
\typeout{** WARNING: IEEEtran.bst: No hyphenation pattern has been}%
\typeout{** loaded for the language `#1'. Using the pattern for}%
\typeout{** the default language instead.}%
\else
\language=\csname l@#1\endcsname
\fi
#2}}

\bibitem{Landauer:94a}
R.~Landauer, ``Is quantum mechanically coherent computation useful?'' in
  \emph{Proc. of the Drexel-4 Symposium on Quantum Nonintegrability}.\hskip 1em
  plus 0.5em minus 0.4em\relax International Press, 1994.

\bibitem{Landauer:95a}
------, ``Is quantum mechanics useful?'' \emph{Philosophical Transactions:
  Physical Sciences and Engineering}, vol. 353, pp. 367--376, 1995.

\bibitem{Chuang:95b}
I.~L. Chuang, R.~Laflamme, P.~W. Shor, and W.~H. Zurek, ``Quantum computers,
  factoring, and decoherence,'' \emph{Science}, vol. 280, pp. 1633--1635, 1995,
  {\tt arXiv:quant-ph/9503007}.

\bibitem{Unruh:95a}
W.~G. Unruh, ``Maintaining coherence in quantum computers,'' \emph{Phys. Rev.
  A}, vol.~51, pp. 992--997, 1995, {\tt arxiv:hep-th/9406058}.

\bibitem{Shor:94a}
P.~W. Shor, ``Algorithms for quantum computation: Discrete log and factoring,''
  in \emph{Proceedings of the 35th Annual Symposium on the Foundations of
  Computer Science}, S.~Goldwasser, Ed.\hskip 1em plus 0.5em minus 0.4em\relax
  Los Alamitos, CA: IEEE Computer Society, 1994, pp. 124--134, {\tt
  arXiv:quant-ph/9508027}.

\bibitem{Grover:96a}
L.~Grover, ``A fast quantum mechanical algorithm for database search,'' in
  \emph{Proceedings of the 28th Annual ACM Symposium on the the Theory of
  Computation}.\hskip 1em plus 0.5em minus 0.4em\relax New York: ACM Press,
  1996, pp. 212--219, {\tt arXiv:quant-ph/9605043}.

\bibitem{Aharonov:97a}
D.~Aharonov and M.~Ben-Or, ``Fault-tolerant quantum computation with constant
  error rate,'' in \emph{Proceedings of the twenty-ninth annual ACM symposium
  on Theory of computing}.\hskip 1em plus 0.5em minus 0.4em\relax ACM Press,
  1997, pp. 176--188, {\tt arXiv:quant-ph/9611025}.

\bibitem{Aharanov:99a}
------, ``Fault-tolerant quantum computation with constant error rate,'' 1999,
  {\tt arXiv:quant-ph/9906129}.

\bibitem{Knill:98a}
E.~Knill, R.~Laflamme, and W.~H. Zurek, ``Resilent quantum computation,''
  \emph{Science}, vol. 279, pp. 342--345, 1998.

\bibitem{Knill:98b}
------, ``Resilient quantum computation: error models and thresholds,''
  \emph{Proc. Roy. Soc. London Ser. A}, vol. 454, pp. 365--384, 1998, {\tt
  arXiv:quant-ph/9702058}.

\bibitem{Kitaev:97b}
A.~Kitaev, ``Quantum error correction with imperfect gates,'' in \emph{Quantum
  Communication, Computing and Measurement}.\hskip 1em plus 0.5em minus
  0.4em\relax New York: Plenum Press, 1997, pp. 181--188.

\bibitem{Kitaev:97c}
------, ``Fault-tolerant quantum computation by anyons,'' \emph{Ann. of Phys.},
  vol. 303, pp. 2--30, 2003, {\tt arXiv:quant-ph/9707021}.

\bibitem{Shor:95a}
P.~W. Shor, ``Scheme for reducing decoherence in quantum memory,'' \emph{Phys.
  Rev. A}, vol.~52, pp. R2493--R2496, 1995.

\bibitem{Steane:96a}
A.~M. Steane, ``Error correcting codes in quantum theory,'' \emph{Phys. Rev.
  Lett.}, vol.~77, p. 893, 1996.

\bibitem{Knill:00a}
E.~Knill, R.~Laflamme, and L.~Viola, ``Theory of quantum error correction for
  general noise,'' \emph{Phys. Rev. Lett.}, vol.~84, pp. 2525--2528, 2000, {\tt
  arXiv.org:quant-ph/9908066}.

\bibitem{Kribs:05a}
D.~Kribs, R.~Laflamme, and D.~Poulin, ``A unified and generalized approach to
  quantum error correction,'' \emph{Phys. Rev. Lett.}, vol.~94, p. 180501,
  2005, {\tt arXiv:quant-ph/0412076}.

\bibitem{Kribs:05b}
D.~W. Kribs, R.~Laflamme, D.~Poulin, and M.~Lesosky, ``Operator quantum error
  correction,'' \emph{Quantum Information \& Computation}, vol.~6, pp.
  382--399, 2005, {\tt arXiv:quant-ph/0504189}.

\bibitem{Kribs:06a}
D.~W. Kribs and R.~W. Spekkens, ``Quantum error correcting subsystems as
  unitarily recoverable subsystems,'' 2006, {\tt arXiv:quant-ph/0608045}.

\bibitem{Nielsen:05a}
M.~A. Nielsen and D.~Poulin, ``Algebraic and information-theoretic conditions
  for operator quantum error-correction,'' 2005, {\tt arxiv:quant-ph/0506069}.

\bibitem{Poulin:05a}
D.~Poulin, ``Stabilizer formalism for operator quantum error correction,''
  \emph{Phys. Rev. Lett.}, vol.~95, p. 230504, 2005, {\tt
  arXiv:quant-ph/0508131}.

\bibitem{Bacon:06a}
D.~Bacon, ``Operator quantum error-correcting subsystems for self-correcting
  quantum memories,'' \emph{Phys. Rev. A}, vol.~73, p. 012340, 2006, {\tt
  arXiv:quant-ph/0506023}.

\bibitem{Klappenecker:06a}
A.~Klappenecker and P.~K. Sarvepalli, ``Clifford code constructions of operator
  quantum error correcting codes,'' 2006, {\tt arXiv.org:quant-ph/0604161}.

\bibitem{Thaker:06a}
D.~D. Thaker, T.~S. Metodi, A.~W. Cross, I.~L. Chuang, and F.~T. Chong,
  ``Quantum memory hierarchies: Efficient designs to match available
  parallelism in quantum computing,'' in \emph{ISCA '06: Proceedings of the
  33rd International Symposium on Computer Architecture}.\hskip 1em plus 0.5em
  minus 0.4em\relax Washington, DC, USA: IEEE Computer Society, 2006, pp.
  378--390, {\tt arXiv:quant-ph/0604070}.

\bibitem{Aliferis:06a}
P.~Aliferis and A.~W. Cross, ``Sub-system fault tolerance with the {Bacon-Shor}
  code,'' 2006, {\tt arXiv:quant-ph/0610063}.

\bibitem{Nielsen:00a}
M.~A. Nielsen and I.~L. Chuang, \emph{Quantum Computation and Quatum
  Information}.\hskip 1em plus 0.5em minus 0.4em\relax New York: Cambridge
  University Press, 2000.

\bibitem{Kempe:01a}
J.~Kempe, D.~Bacon, D.~A. Lidar, and K.~B. Whaley, ``Theory of decoherence-free
  fault-tolerant quantum computation,'' \emph{Phys. Rev. A}, vol.~63, pp.
  042\,307--1--042\,307--29, 2001, {\tt arXiv:quant-ph/0004064}.

\bibitem{Zanardi:01a}
P.~Zanardi, ``Stabilizing quantum information,'' \emph{Phys. Rev. A}, vol.~63,
  p. 12301, 2001, {\tt arXiv:quant-ph/9910016}.

\bibitem{Bartlett:03a}
S.~D. Bartlett, T.~Rudolph, and R.~W. Spekkens, ``Classical and quantum
  communication without a shared reference frame,'' \emph{Phys. Rev. Lett.},
  vol.~91, p. 027901, 2003, {\tt arXiv:quant-ph/0302111}.

\bibitem{Bacon:04a}
D.~Bacon, I.~L. Chuang, and A.~W. Harrow, ``Efficient quantum circuits for the
  {Schur and Clebsch-Gordan} transforms,'' 2004, {\tt arXiv:quant-ph/0407082}.

\bibitem{Bacon:06b}
------, ``The quantum {S}chur transform: I. efficient qudit circuits,'' 2006,
  {\tt arXiv:quant-ph/0601001}.

\bibitem{Steane:98a}
A.~M. Steane, ``Space, time, parallelism and noise requirements for reliable
  quantum computing,'' \emph{Fortsch. Phys.}, vol.~46, pp. 443–--458, 1998,
  {\tt arXiv:quant-ph/9708022}.

\end{thebibliography}

\end{document}